\documentstyle[12pt]{article}
\def \be {\begin{equation}}
\def \ee {\end{equation}}
\begin{document}

\thispagestyle{empty}
\setcounter{page}0

{\flushright
JINR preprint E2-94-253 \\
Dubna, July 5, 1994\\}
\vfill

\begin{center}
{\Large\bf New Tensor Interactions \\
{}~\\
and the $K_L$-$K_S$ Mass Difference}

\vfill

{\large M. V. Chizhov}\footnote{E-mail: $physfac2@bgearn.bitnet$}
 \vspace{1cm}

{\em Bogoliubov Laboratory of Theoretical Physics, Joint Institute for
Nuclear Research, Dubna, Russia}\\ and \\ {\em
Center of Space Research and Technologies, Faculty of Physics,
University of Sofia, 1126 Sofia, Bulgaria}\footnote{Permanent address}

\end{center}

\vfill

\begin{abstract}
The minimal standard electroweak model yields approximately
half the observed value of the $K_L$-$K_S$ mass difference.
Antisymmetric tensor fields incorporated into the standard
model can provide the rest $0.5\times(\Delta m_{LS})_{\exp}$.
The effective lepton--lepton, quark--lepton and quark--quark
tensor interactions induced by the charged tensor particles
are presented.
\end{abstract}

\vfill

{\em Submitted to ``Phys. Lett. B''}

\vfill

\newpage

\section{Introduction} \indent

The triumph of the standard electroweak model does not leave much place
for new physics when describing heavy quark interactions, still less for
the description of old hadron weak decays. Most of the papers revising
the calculations in the $\pi$- and $K$-meson physics try to find more
accurate solutions of the problems connected with the strong
interactions, but they do not change the base of the weak V--A
interactions. However, the recent experimental results for the radiative
$\pi^- \to e^- \bar{\nu} \gamma$ decay in a wide range of
kinematic variables \cite{Bol} and the three-particle $K^+ \to
\pi^0 e^+ \nu$ decay \cite{Ak} (see also Ref.\cite{St}) point to the
impossibility of their adequate description in the framework of the V--A
model. In order to explain the destructive interference in the pion
decay, in Ref.\cite{Pob1} an additional tensor quark--lepton interaction
was suggested
\be
{\cal{L}}_T = \sqrt 2~ G_F V_{ud} f_T\;\bar{u}_R \sigma_{\mu\nu} d_L \cdot
\bar{e}_R \sigma^{\mu\nu} \nu_L,
\ee
where $V_{ud}$ is an element of the Kobayashi--Maskawa mixing matrix and
$\sigma_{\mu\nu} = {1 \over 2}i\; [\gamma_{\mu}, \gamma_{\nu}]$. This
interaction provides consistency with other experimental data on this
decay \cite{Pob2}. The dimensionless constant $f_T$ describes the
strength of the interaction relative to the ordinary weak coupling. The
constant $f_T$ is estimated in the framework of the relativistic quark
model as $f^{\rm RQM}_T= -(4.2\pm 1.3)\times 10^{-2}$. In
Ref.\cite{Bel}, calculations of $f_T$ have been done by applying the QCD
techniques and the PCAC hypothesis. The value obtained in this way
$f^{\rm QCD}_T=-(1.4\pm 0.4)\times 10^{-2}$ is one third as low as
estimated in the framework of the relativistic quark model. It seems
that the accuracy of the calculations in Ref.\cite{Bel} is the usual
PCAC accuracy and that the latter result is reliable.

As was noted in Ref.\cite{Vol}, such a naive form
of interaction will lead, as a result of radiative corrections, to
an anomalously
large contribution to the $\pi \to e\nu$ decay, in contradiction with
the experimental data. To avoid this difficulty and to describe all the
meson-decay experiments [1--3] simultaneously, in Ref.\cite{Chiz} it was
suggested to extend the standard model by adding two doublets of new
tensor particles. Thus extended electroweak model predicts, besides the
appearance of a new quark--lepton interaction like Eq.(1), also the
appearance of tensor interactions in the lepton--lepton and
quark--quark sectors.

The tensor interaction does not conserve chirality, and therefore, it
can play an important role in the nonleptonic weak processes.
Particularly, in this paper we are going to demonstrate that the
effective tensor quark--quark interaction gives a contribution to the
$K_L$-$K_S$ mass difference $\Delta m_{LS}$ of the order of the
contribution from the standard V--A interaction. This is easy to
understand because the strength of the new interaction is just an order
of magnitude as low as $G_F$. On the other hand, from the
PCAC hypothesis it follows that the chirality-nonconserving currents are
enhanced by a factor $\chi=(m_K/m_s)^2 \sim 10$, where $m_K$
and $m_s$ are the masses of the kaon and the strange quark,
respectively. It is known that the minimal standard V--A model
(without taking into account large-distance contributions) can explain
only half the experimental value of the mass difference $(\Delta
m_{LS})_{\exp}$ \cite{Shif}.  Formally, the large-distance contribution
is suppressed by $\mu^2/m^2_c$, where $\mu$ is a characteristic mass
scale in the ``old" hadron physics. Neglecting the large-distance
contribution for the neutral kaon system, we suppose that the remaining
half of $(\Delta m_{LS})_{\exp}$ is due to the new tensor interactions
of quarks. This completely fixes the parameters of the new effective
tensor interactions in all sectors. The effect of the
chirality-nonconserving currents in the lepton--lepton sector (for
example, in $\mu \to e \bar{\nu}_e \nu_\mu$ decay) is suppressed by the
factor $m_e/m_\mu$ and falls into the experimental accuracy range.  The
new tensor interaction in the lepton--quark sector does not contradict
the existing constraints from nuclear beta-decay \cite{Pob1} either.

\section{New effective tensor interactions} \indent

The antisymmetric second-rank tensor fields are incorporated into the
standard electroweak model as two doublets \cite{Chiz}
$T_{\mu\nu}=(T^+_{\mu\nu}, T^0_{\mu\nu})$ and $U_{\mu\nu}=(U^0_{\mu\nu},
U^-_{\mu\nu})$ with opposite hypercharges: $Y(T)=-Y(U)=+1$. The Yukawa
interaction of the charged components of the tensor fields with the
first generation of leptons $\nu_e$, $e$ and the quarks $u$, $d'=d
\cos\theta_C + s \sin\theta_C$ is
\be
{\cal{L}}_{\rm Yukawa} = {t\over 2} \left[ \bar{\nu}_L\sigma^{\mu\nu}e_R
 + \bar{u}_L\sigma^{\mu\nu} d_R' \right] \cdot T^+_{\mu\nu} +
{t\over 2} \bar{u}_R\sigma^{\mu\nu} d_L'\cdot U^+_{\mu\nu} + {\rm h.c.}
\ee
For simplicity we accept the coupling constants for quarks and leptons
to have the same value $t$. After the spontaneous symmetry breaking, a
mixing of the $T$ and $U$ fields occurs. The propagator of the charged
tensor fields in the static approximation $q^2 \ll m^2,M^2$ is of the
form
\begin{eqnarray}
{\cal{P}}(q) &=& \left(
 \begin{array}{cc}
  <T(T^+T^-)>_0 & <T(T^+U^-)>_0 \\ <T(U^+T^-)>_0 & <T(U^+U^-)>_0
 \end{array} \right) \nonumber\\
&=& {2i \over m^2-M^2} \left(
 \begin{array}{cc}
  {\bf \Pi}(q) & -{\bf 1} \\ -{\bf 1} & \tan^2\beta~ {\bf \Pi}(q)
 \end{array} \right),
\end{eqnarray}
where ${\bf 1}_{\mu\nu\alpha\beta}={1\over2} \left(
g_{\mu\alpha}~g_{\nu\beta} - g_{\mu\beta}~g_{\nu\alpha} \right)$,
$$
{\bf \Pi}_{\mu\nu\alpha\beta} (q) = {\bf 1}_{\mu\nu\alpha\beta} -
{q_{\mu}q_{\alpha} g_{\nu\beta} - q_{\mu} q_{\beta} g_{\nu\alpha}
 - q_{\nu} q_{\alpha}g_{\mu\beta} + q_{\nu} q_{\beta} g_{\mu\alpha}
 \over q^2} ,
$$
and $\tan\beta=M/m$ is the ratio of the two mass parameters, associated
with the vacuum expectation values of the two doublets of the Higgs
particles. The diagonalization of Eq.(3) gives the mass matrix of the
form ${\cal{M}}^2=M^2 {\rm diag}(\lambda_T, \lambda_U)$ for the fields
\be
\begin{array}{l}
T_{\mu\nu}'= T_{\mu\nu} \cos\varphi + \Pi_{\mu\nu\alpha\beta}
 U^{\alpha\beta}\sin\varphi, \\
U_{\mu\nu}'= -\Pi_{\mu\nu\alpha\beta} T^{\alpha\beta} \sin\varphi +
 U_{\mu\nu}\cos\varphi,
\end{array}
\ee
where
$$\lambda_T={1\over 2}
 \left[ 1+\tan^2\beta+\sqrt{(1-\tan^2\beta)^2+4}
 \right] /\tan^2\beta,$$
$$\lambda_U={1\over 2}
 \left[ 1+\tan^2\beta-\sqrt{(1-\tan^2\beta)^2+4}
 \right] /\tan^2\beta, $$
and $\tan\varphi={1\over 2} \left[ 1 -\tan^2\beta +\sqrt{
(1-\tan^2\beta)^2 +4 } \right] $. The positive definiteness of the
matrix of squared masses leads to the restriction $\tan^2\beta>1$. The
dependence of the squared masses of the fields $T'$ and $U'$ on the
parameter $\cot^2\beta$ is shown in Fig.1.

Using lagrangian (2) and propagator (3), one can easily write the
effective interactions for various sectors. For example, in the case of
a muon decaying into an electron and (anti)neutrinos, one must add to
the usual V--A interaction an interaction of the form
\be
{\cal{L}}_{\mu e}
 = - \sqrt 2~ G_F f_t\; \bar{\nu}_{\mu L} \sigma_{\alpha\lambda} \mu_R
{}~{q^{\alpha}q_{\beta} \over q^2}~
\bar{e}_R \sigma^{\beta\lambda}\nu_{eL}
+ {\rm h.c.},
\ee
where $q$ is the momentum transferred from the muon to the electron
pair, and the constant $f_t$ is defined as $f_t~G_F/\sqrt 2 = t^2
/(M^2-m^2)$. The additional interaction for semileptonic weak decays has
the form
\be
{\cal{L}}_{qe} = - \sqrt 2~ G_F f_t\; \bar{u} \sigma_{\mu\lambda} d'
 ~{q^{\mu}q_{\nu} \over q^2}~ \bar{e}_R \sigma^{\nu\lambda}\nu_{eL}
 + {\rm h.c.}
\ee
The constant $f_t$ can be fixed from the decays $\pi^- \to e^- \bar{\nu}
\gamma$ or $K^+ \to \pi^0 e^+ \nu$. Its value $f_t^{\rm RQM}=0.29\pm
0.09$ \cite{Chiz} was evaluated in the framework of the relativistic
quark model on the basis of the pion-decay experiment \cite{Bol}. In
what follows we are going to work in the framework of QCD and
extensively use the PCAC hypothesis. Therefore, we shall take the
corrected value $f_t^{\rm QCD}=(7.84\pm 2.24)\times 10^{-2}$ in
accordance with Ref.\cite{Bel}.

A richer interaction is obtained in the pure quark--quark sector:
\begin{eqnarray}
{\cal{L}}_{ud} &=& - \sqrt 2~ G_F f_t \;
\left[ \bar{u}_L \sigma_{\mu\lambda} d_R' \cdot
 \bar{d}_R' \sigma^{\nu\lambda} u_L +
 \bar{u}_L \sigma_{\mu\lambda} d_R' \cdot
 \bar{d}_L' \sigma^{\nu\lambda} u_R
\right. \nonumber \\*
&&
\left.
 + \bar{u}_R \sigma_{\mu\lambda} d_L' \cdot
  \bar{d}_R' \sigma^{\nu\lambda} u_L
 + \tan^2\beta~ \bar{u}_R \sigma_{\mu\lambda} d_L' \cdot
  \bar{d}_L' \sigma^{\nu\lambda} u_R
\right] {q^{\mu}q_{\nu} \over q^2}.
\end{eqnarray}
The independent unknown parameter $\tan\beta$ appears just in the last
term of the quark--quark interaction (7). Therefore, in order to fix it,
it is natural to investigate some weak nonleptonic process. The
$K_L$-$K_S$ mass difference is most sensitive to new hypothetical
particles and interactions. We will prove that the introduced new
interaction (7) does not contradict $(\Delta m_{LS})_{\exp}$.

\section{The kaon mass difference} \indent

In the $K^0$-$\bar{K}^0$ mixing calculations we shall take into account
only the contributions of the $u$ and $c$ quarks (see the box diagrams
of Fig.2), because the $t$-quark contribution is suppressed by the tiny
mixing angle $V_{td}$. As the standard contribution that results from
the $W$-boson exchange (Fig.1a) gives only half the experimental value
$(\Delta m_{LS})_{\exp}$, we suppose that the rest of $(\Delta
m_{LS})_{\exp}$ is due to the other two diagrams with the exchange of
the tensor particles (Fig.2b,c). The effective lagrangian for the second
generation of quarks, caused by the tensor particles, has the same form
as Eq.(7) with obvious substitutions $u \to c$ and $d' \to
s'=-d\sin\theta_C+s\cos\theta_C$:
\begin{eqnarray}
{\cal{L}}_{cs} &=& - \sqrt 2~ G_F f_t \;
\left[
 \bar{c}_L \sigma_{\mu\lambda} s_R' \cdot
  \bar{s}_R' \sigma^{\nu\lambda} c_L
 + \bar{c}_L \sigma_{\mu\lambda} s_R' \cdot
  \bar{s}_L' \sigma^{\nu\lambda} c_R
\right. \nonumber \\*
&&
\left.
 + \bar{c}_R \sigma_{\mu\lambda} s_L' \cdot
  \bar{s}_R' \sigma^{\nu\lambda} c_L
 + \tan^2\beta~ \bar{c}_R \sigma_{\mu\lambda} s_L' \cdot
  \bar{s}_L' \sigma^{\nu\lambda} c_R
\right] {q^{\mu}q_{\nu} \over q^2}.
\end{eqnarray}
Only the first terms in Eqs.(7) and (8) contribute to the interference
diagrams (Fig.2b)
\be
{\cal{L}}_{WT}
 = - {3 G^2_F f_t \over 16 \pi^2} \sin^2\theta_C \cos^2\theta_C \; m^2_c
{}~\bar{s}(1-\gamma^5)d \cdot \bar{s}(1+\gamma^5)d.
\ee
The divergency is eliminated by the GIM mechanism, as in diagrams of
Fig.2a. The matrix element
\be
<\bar{K}^0 \mid \bar{s}(1-\gamma^5)d \cdot \bar{s}(1+\gamma^5)d \mid K^0>
= - \left( 2\chi +{1 \over 3} \right) F^2_K m^2_K
\ee
is calculated in the approximation of the vacuum saturation, as the
vacuum is inserted in all possible channels. Assuming $F_K$=160~MeV and
$m_c$=1.35~GeV, we can estimate the contribution of the interference
diagrams to $\Delta m_{LS}={\rm Re}<\bar{K}^0 \mid {\cal{L}}_{\Delta
S=2} \mid K^0>/m_K$ as $1.3\times(\Delta m_{LS})_{\exp}$.

Although the contribution of the diagrams in Fig.2c seems of the order
of the small parameter $f^2_t$, actually, their contribution is enhanced
because the suppressing GIM mechanism does not work. Therefore, they
must be taken into account. We assume that the masses of the new tensor
particles are of the order of the vector-boson mass $M_W$. Then the
logarithmically divergent integral is effectively cut off at this limit.
Such an assumption will not essentially change our results, as the
logarithm is a very slowly increasing function. The effective lagrangian
for the $K^0$-$\bar{K}^0$ mixing due to exchange of the tensor particles
(Fig.2c) reads
\begin{eqnarray}
 {\cal{L}}_{TT} &=& - {3 G^2_F f^2_t \over 16 \pi^2}
 \sin^2\theta_C \cos^2\theta_C \; m^2_c
\left\{ -{1 \over 4} (\bar{s}_R \gamma_\mu d_R)^2
 + {3 \over 2} (\bar{s}_R \gamma_\mu d_R)~ (\bar{s}_L \gamma^\mu d_L)
\right. \nonumber\\*
&&
\left. - {\tan^4\beta \over 4} (\bar{s}_L \gamma_\mu d_L)^2
 - \ln{M^2_W \over m^2_c}
 \left[ (\bar{s}_R d_L)^2 + {1 \over 6}(\bar{s}_R \sigma_{\mu\nu} d_L)^2
 \right.
\right. \nonumber\\*
&&
\left.
 \left. +2 \tan^2\beta~ (\bar{s}_R d_L)(\bar{s}_L d_R) +
  (\bar{s}_L d_R)^2 + {1 \over 6}(\bar{s}_L \sigma_{\mu\nu} d_R)^2
 \right]
\right\} .
\end{eqnarray}
The matrix elements
\be
<\bar{K}^0 \mid \bar{s} \gamma_\mu (1+\gamma^5) d \cdot
\bar{s} \gamma^\mu (1-\gamma^5) d \mid K^0> =
(2+{4 \over 3}\chi) F^2_K m^2_K,
\ee
\be
<\bar{K}^0 \mid \left[\bar{s} \gamma_\mu (1 \pm \gamma^5) d
\right]^2 \mid K^0> =
- {8 \over 3} F^2_K m^2_K,
\ee
\be
<\bar{K}^0 \mid \left[\bar{s} (1 \pm \gamma^5) d
\right]^2 + {1 \over 6}\left[\bar{s} \sigma_{\mu\nu} (1 \pm \gamma^5) d
\right]^2 \mid K^0> =
\chi F^2_K m^2_K,
\ee
are calculated as before with the aid of the vacuum saturation
hypothesis. To give the correct total mass difference $(\Delta
m_{LS})_{\exp}$, the contribution of the diagrams in Fig.2c must be
negative $-0.8\times(\Delta m_{LS})_{\exp}$. This is really the case: in
the whole allowed range of the parameter $\tan^2\beta>1$, the
contribution of Fig.2c has the opposite sign relative to Figs.2a,b.

The correct value of the mass difference is obtained when $\tan\beta =
1.51\pm 0.07$. Notice that this solution
$\cot^2\beta=0.44\pm 0.04$ is very close to the value 0.4 that
corresponds to the maximal possible mass of the tensor field
$U_{\mu\nu}$ (see Fig.1). That would provide the minimal effective
potential between the fermions. It is curious to note that the
mixing angle $\varphi$ happens to be surprisingly close to the Weinberg
angle $\theta_W$:  $\sin^2\varphi=0.229\pm 0.027$.

Now let us discuss the effect of the two Higgs doublets on the mass
difference $(\Delta m_{LS})_{\exp}$. Their Yukawa coupling constants
with quarks are proportional to the quark masses. As far as the
model-independent lower bound on the $t$-quark mass is $m_t$$>$90~GeV
(i.e. the $t$ quark is very heavy), the contribution of the virtual $t$
quarks to the box diagrams may become dominant in spite of the strong
suppression of the Yukawa interaction owing to the tiny mixing angle
$V_{td}$. If the ratio of the vacuum expectation values of the two
doublets of the Higgs particles $v_1/v_2= M/m=\tan\beta$ were large,
this could enhance the contribution of the Higgs particles to the
$K_L$-$K_S$ mass difference (according to the analysis in
Ref.\cite{Abb}). However, the value of $\tan\beta \approx 1.51$
estimated above does not lead to a large ratio $v_1/v_2$, and
respectively, gives no considerable enhancement of the contribution of
the two Higgs doublets to the $K_L$-$K_S$ mass difference as compared to
the case of only one doublet in the minimal standard model.

There exists a great uncertainty in the values of the Kobayashi--Maskawa
matrix elements for the $t$ quark. We use here the central values
$V_{td}=0.010$ and $V_{ts}=0.042$ \cite{Data}. In Fig.3 the relative
contribution of the Higgs doublets to the actual $K_L$-$K_S$ mass
difference is shown. The curves in the plane of two parameters --- the
charged-Higgs mass $M_H$ and the $t$-quark mass $m_t$ --- correspond to
the fixed values of the relative contribution. Obviously, the
contribution of the Higgs doublets increases with the increase of the
$t$-quark mass and the decrease of the Higgs mass. There exists a bound
on the value of the $t$-quark mass from the Tevatron: $m_t=174\pm
10^{+13}_{-12}$GeV \cite{Ferm}. On the other hand, it is natural to
assume that the tensor particles are included into a more complete
supersymmetric theory which ensures the unification of the coupling
constants and of the $b$--$\tau$ masses \cite{Kol}. Then, using the
value of $\tan\beta = 1.51\pm 0.07$ obtained above, one can estimate the
$t$-quark mass as $m_t=165\pm 13$~GeV. In these theories the masses of
the charged Higgs particles are above $260$~GeV. Therefore, for the
realistic choice $m_t$=170~GeV and $M_H$$>$260~GeV, the contribution to
$\Delta m_{LS}$ is less than $10\%$. Such an accuracy is compatible
with the assumptions made: the PCAC hypothesis, the vacuum-saturation
approximation, disregard of the large-distance contributions, etc.

In interactions of $up$ and $down$ quarks with the charged tensor
particles, two additional generation-mixing matrices of the
Cabibbo--Kobayashi--Maskawa type arise: between left--right and
right--left quarks. Unfortunately, because of the lack of information
about these matrices, we cannot say anything more definite about the
contribution of the tensor particles to the $CP$ violation in the
$K^0$-$\bar{K}^0$ system.

\section{Conclusions} \indent

The experimental data [1--3] point to an admixture of tensor currents in
the weak decays. In the electroweak model extended by the doublets of
tensor particles, we can fix all the parameters of effective
interactions of the charged tensor currents, using the cited experiments
and the $K_L$-$K_S$ mass difference. The model predicts neutral tensor
currents too, but it is as yet impossible to define this sector because
of the lack of experimental data. At the moment we can definitely say
only that there exists a requirement that the flavor-changing neutral
tensor currents should be absent.

When estimating the mixing parameter $\tan\beta$ from the $K_L$-$K_S$
mass difference, we have neglected the QCD corrections. Taking them
into account may correct the value of $\tan\beta$ --- the presented
value $\tan\beta \approx 1.51$ was obtained for bare quarks. As we have
accepted the same coupling of the tensor fields to leptons and quarks,
the effective lepton--lepton and quark--lepton tensor interactions are
governed by a single common parameter $f_t$. Therefore, by fixing this
parameter from the semileptonic decays, $f_t \approx 0.08$, we can make
a prediction for the weak leptonic decays, where no uncertainty
associated with the strong interactions is present.

Unfortunately, all previous attempts to study the violation of the V--A
structure of interaction in the $\mu \to e \bar{\nu}_e \nu_\mu$ decay
have been done on the basis of the derivative-free effective lagrangian
\cite{Weak}. The effective tensor interactions presented here
essentially depend on the momentum transfer. Certainly, this will lead
to a different angular and energy distribution of the outgoing electron.
Nevertheless, we can estimate the contribution of the new interaction to
the Michel parameter $\rho$. Its value $\delta\rho \sim 10^{-3}$ is of
the same order of magnitude as the present experimental accuracy
\cite{Data}. \vspace{5mm}

\pagebreak[3]
{\bf Acknowledgements} \vspace{3mm}

I am grateful to L.V.Avdeev, D.I.Kazakov and M.D.Mateev for helpful
discussions. I acknowledge the hospitality of the Bogoliubov Laboratory
of Theoretical Physics, JINR, Dubna, where this work has been completed.
The work is financially supported by Grant-in-Aid for Scientific
Research F-214/2096 from the Bulgarian Ministry of Education, Science
and Culture.

\pagebreak[4]

\newpage
\setlength\unitlength{4pt}
\pagestyle{empty}

{}~\vspace{19cm}
\thicklines
\begin{figure}[htbp]

\caption{The dependence of the squared masses of the $T$ and $U$
particles on the mixing parameter $\cot^2\beta$}

\end{figure}

\vfill

M. V. Chizhov

{\em Phys. Lett. B}

\newpage

\begin{figure}[htbp]
$$\begin{array}{cc}
 \begin{picture}(50,45)(0,-30)
  \multiput(11,10)(4,0)5{\oval(2,2)[t]}
  \multiput(13,10)(4,0)4{\oval(2,2)[b]}
  \multiput(10,10)(18,0)2{\line(0,-1){18}}
  \put(10,1){\vector(0,1)1}
  \put(28,1){\vector(0,-1)1}
  \multiput(11,-8)(4,0)5{\oval(2,2)[b]}
  \multiput(13,-8)(4,0)4{\oval(2,2)[t]}
  \multiput(10,10)(0,-18)2{\line(-1,0){10}}
  \multiput(28,10)(0,-18)2{\line(1,0){10}}
  \multiput(5,10)(28,0)2{\vector(-1,0)1}
  \multiput(5,-8)(28,0)2{\vector(1,0)1}
  \multiput(18,12)(0,-18)2{$W$}
  \put(1,11){$\overline{s}$}
  \put(35,11){$d$}
  \put(1,-7){$d$}
  \put(35,-7){$\overline{s}$}
  \multiput(3,0)(27,0)2{$u,c$}
  \put(18,-15){(a)}
 \end{picture}
 &
 \begin{picture}(50,45)(0,-30)
  \multiput(11,10)(4,0)5{\oval(2,2)[t]}
  \multiput(13,10)(4,0)4{\oval(2,2)[b]}
  \multiput(10,10)(18,0)2{\line(0,-1){17.5}}
  \put(10,1){\vector(0,1)1}
  \put(28,1){\vector(0,-1)1}
  \multiput(10,-7.5)(0,-1)2{\line(1,0){18}}
  \multiput(10,10)(28,0)2{\line(-1,0){10}}
  \multiput(9.5,-8)(28.5,0)2{\line(-1,0){9.5}}
  \multiput(5,10)(28,0)2{\vector(-1,0)1}
  \multiput(5,-8)(28,0)2{\vector(1,0)1}
  \multiput(10,-8)(18,0)2{\circle 1}
  \put(18,12){$W$}
  \put(18,-6){$T$}
  \put(1,11){$\overline{s}$}
  \put(35,11){$d$}
  \put(1,-7){$d$}
  \put(35,-7){$\overline{s}$}
  \multiput(3,0)(27,0)2{$u,c$}
  \put(18,-15){(b)}
 \end{picture}
 \\
 &
 \begin{picture}(50,45)(0,-30)
  \multiput(10,9.5)(0,1)2{\line(1,0){18}}
  \multiput(10,9.5)(18,0)2{\line(0,-1){17}}
  \put(10,1){\vector(0,1)1}
  \put(28,1){\vector(0,-1)1}
  \multiput(10,-7.5)(0,-1)2{\line(1,0){18}}
  \multiput(9.5,10)(28.5,0)2{\line(-1,0){9.5}}
  \multiput(9.5,-8)(28.5,0)2{\line(-1,0){9.5}}
  \multiput(5,10)(28,0)2{\vector(-1,0)1}
  \multiput(5,-8)(28,0)2{\vector(1,0)1}
  \multiput(10,-8)(18,0)2{\circle 1}
  \multiput(10,10)(18,0)2{\circle 1}
  \put(18,12){$T$}
  \put(18,-6){$T$}
  \put(1,11){$\overline{s}$}
  \put(35,11){$d$}
  \put(1,-7){$d$}
  \put(35,-7){$\overline{s}$}
  \multiput(3,0)(27,0)2{$u,c$}
  \put(18,-15){(c)}
 \end{picture}
\end{array}$$

 \caption{Box diagrams for $K_L$-$K_S$ mass difference}
\end{figure}

\vfill

M. V. Chizhov

{\em Phys. Lett. B}

\newpage

{}~\vspace{19cm}

\begin{figure}[htbp]

\caption{Isolines of the relative contribution from the Higgs doublets
to the $K_L$-$K_S$ mass difference in the plane of the charged-Higgs
mass and the $t$-quark mass}

\end{figure}

\vfill

M. V. Chizhov

{\em Phys. Lett. B}

\end{document}